\documentclass[a4paper,12pt]{article}
\usepackage[utf8]{inputenc}
\usepackage{amsfonts,mathrsfs,amsmath}
\usepackage{authblk,bm}
\usepackage{natbib,graphicx}
\begin{document}
\title{Influence of the black hole spin on the chaotic particle dynamics within a dipolar halo}
\author[a]{Sankhasubhra Nag\thanks{email: sankha@sncwgs.ac.in}}
\author[b]{Siddhartha Sinha}
\author[c]{Deepika B. Ananda\thanks{Present Address: Nicolaus Copernicus Astronomical Centre of the Polish Academy of Sciences, Warsaw; email: deepika@camk.edu.pl}} 
\author[c]{\\ Tapas K Das\thanks{email: tapas@hri.res.in}}
\affil[a]{\small Sarojini Naidu College for Women, Kolkata 700 028, India}
\affil[b]{\small Khatra Adibasi Mahavidyalaya, Khatra 722140, India}
%\affil[c]{\small Indian Institute of Science Education and Research, Pune 411 008, India}
\affil[c]{\small Harish-Chandra Research Institute, Allahabad 211 019, India}
\date{}
\maketitle
\begin{abstract}
\noindent
We investigate the role of the spin angular momentum of astrophysical black holes 
in controlling the special relativistic chaotic dynamics of test particles moving under the 
influence of a post-Newtonian pseudo-Kerr black hole potential, along with a 
perturbative potential created by a asymmetrically placed (dipolar) halo.
Proposing a Lyapunov-like exponent to be the effective measure of the degree of
chaos observed in the system under consideration, 
it has been found that black hole spin anti-correlates with the degree of chaos 
for the aforementioned dynamics. Our findings have
been explained applying the general principles of dynamical systems analysis.
\end{abstract}

\section{Introduction}
\noindent
Investigation of the chaotic dynamics of test particles within accretion discs 
or inside the halo surrounding astrophysical black holes have gained widespread 
interest in recent years. Integrability  conditions and the 
possibility of transition to the chaotic motion for charged particles 
moving under the influence of the magnetic force and strong gravitational fields of compact 
objects and its magnetospheres~\citep{Kopacek2010,Kopacek2014,Kopacek2015,Takahashi2009, Kovar2008,Kovar2010},
motion under the influence of the 
relativistic gravitational field of an accreting black hole systems~\citep{Semrak2010,Semrak2012,Semrak2013,Witzany2015,Vogt2003},
and motion under the influence of gravity produced by an extended body~\citep{Letelier1997,Castro2011},
have been studied within the general relativistic framework. 
Post-Newtonian black hole potentials have been used to study the chaotic motion 
of particles moving inside the halo surrounding a non-rotating (\cite{Gueron2001,Gueron2001a,Letelier2011,Chen2003})
or a rotating black hole \citep{Ying2012}. 

In the present work we would like to study the chaotic motion of test particles 
under the combined influence of the gravitational field of a Kerr black hole as 
well as the dipolar halo surrounding it by incorporating the potential 
proposed by ~\citet{Artemova1996}. The same pseudo-Kerr black hole potential
has been used by \cite{Ying2012} to study the non 
relativistic (for $v_{\rm particle}<<c$, $c$ being the velocity of 
light in vacuum) chaotic dynamics within the quadruple halo. 

For a halo with symmetric mass distribution, however,
the first non-central leading order term in the potential is quadrapoular,
accommodating both oblate and prolate spheroidal 
geometries with different signs of quadruple parameter~\citep{Ying2012}. For the 
halos with asymmetric mass distribution with respect to the equatorial plane of a spinning black hole the 
dipolar term exists~\citep{Binney2013} and is expected to 
cast stronger effects -- in comparison to the quadrupole term  -- 
in the corresponding dynamics as a perturbing term in general. 

In absence of perturbative effects
produced by the halo, the black hole potential being the central one,
does not leave any room for chaotic motion, nevertheless, as soon as any non-central force, 
however small, comes into the effect -- the system becomes non-integrable~\citep{Goldstein2001,Berry1978}.

We thus believe that in the astrophysical context, 
our formalism presented in this paper is more realistic 
than studying the dynamical system inside a quadrupolar halo. 

As a step forward to existing works studying the non relativistic 
($v_{\rm particle}<<c$) chaotic particle dynamics 
in the pseudo-Schwarzschild/Kerr space-time, we provide a special 
relativistic treatment of the corresponding dynamics, to make our 
formalism more astrophysically relevant and realistic. 
\section{Formulation of the problem and the governing equations}
The free fall acceleration on a Kerr black hole under consideration 
can be mimicked in \citet{Artemova1996} as \[F=-\frac{1}{r^{2-\beta}(r-r_1)^{\beta}},\] where $r_1=1+\sqrt{1-a^2}$ is the radial position of the event horizon in the equatorial plane, for the rotating black hole with effective Kerr parameter $a$. Again,\[ \beta=\frac{r_{in}}{r_1}-1,\] where, \[r_{in}=3+Z_2-\left[(3-Z_1)(3+Z_1+2Z_2)\right]^{\frac{1}{2}},\] with \[Z_1=1+(1-a^2)^{\frac{1}{3}}\left[(1+a)^{\frac{1}{3}}+(1-a)^{\frac{1}{3}}\right] \;\;\textrm{  and  }\;\; Z_2=(3a^2+Z_1^2)^{\frac{1}{2}}.\] By integrating this acceleration and fixing the integration constant so that it asymptotically vanishes at infinite distance, the potential is found to be \begin{equation}                                                                                                                                                                                                                              \Phi(r)= -\frac{1}{r_1(\beta-1)}\left[\frac{r^{\beta-1}}{(r-r_1)^{\beta-1}}-1\right]         ,                                                                                                                                                                                                                   \end{equation}
which is applicable only near the equatorial plane with respect to the black hole spin. Due to the azimuthal symmetry present in the problem, to exploit that, it is better to write it in the cylindrical polar ($\rho, \phi, z$) form  as \begin{equation}
\Phi(\rho, \phi, z)= -\frac{1}{r_1(\beta-1)}\left[\frac{(\rho^2+z^2)^{\frac{\beta-1}{2}}}{(\sqrt{\rho^2+z^2}-r_1)^{\beta-1}}-1\right]  ,                                                                                                                                                                                                                                                                                                                                                                                                                                                                                \end{equation}
where $\phi$ is absent due to the azimuthal symmetry; the equatorial plane is specified by $z=0$.

As mentioned earlier, if the mass distribution of the halo around the black hole is axially symmetric about the spinning axis of the black hole and nearly spherical in shape but yet somewhat displaced about the equatorial plane~\citep{Binney2013}, the gravitational potential part due the effect of halo  will be dominated by leading order dipolar term  of the form $\alpha z$. Hence the net gravitational potential on  a test particle within the halo with the centrally located spinning black hole may be expressed~\citep{Gueron2001,Binney2013} as,  
\begin{equation}
 \Phi_g= \Phi(\rho, \phi, z)+\alpha z,
\end{equation}
where $\alpha$ is a constant, value of which can be set depending upon the asymmetry of the mass distribution in the halo about the equatorial plane.

With respect to the gravitational attraction exerted by the black hole, the effect of halo may be taken as a perturbative contribution 
(with a small, physically realistic value of $\alpha$) making the integrable central force system slightly non-integrable one~\citep{Berry1978}. The angular momentum of the test particle (chosen to be of unit mass for numerical calculations), with respect to the symmetry axis which is taken to be the spinning axis of the black hole too, may be denoted by $L$ and it must be a constant due to the azimuthal symmetry of the problem. Thus the motion along $\phi$ direction can be taken care of by introducing a centrifugal force term along the radial direction, so that the effective potential becomes,\[V(\rho,\phi,z)=-\frac{1}{r_1(\beta-1)}\left[\frac{(\rho^2+z^2)^{\frac{\beta-1}{2}}}{(\sqrt{\rho^2+z^2}-r_1)^{\beta-1}}-1\right]+\alpha z+\frac{L^2}{2\rho^2}.\]

Thus the equations of motion can be expressed as,
\begin{subequations}\begin{align}\dot{\rho} &= p_{\rho},\\ \dot{p}_{\rho} &= -\frac{\partial\Phi_g}{\partial \rho}+\frac{L^2}{\rho^3}=-\frac{\partial V}{\partial\rho},\\ \dot{z} &= p_z, \\ \dot{p}_z &= -\frac{\partial \Phi_g}{\partial z}=-\frac{\partial V}{\partial z};\end{align}\label{govern}\end{subequations} where all the velocities are scaled by the speed of light in vacuum i.e. $c$ and all the distances are scaled by $GM_{\rm BH}/c^2$, $M_{\rm BH}$ being the mass of the centrally located black hole.
                    
As the velocity components may turn out not to be negligible with $c$ (i.e. $=1$ after the scaling), the equations of motion may require the special relativistic correction and after such corrections those may look like,
\begin{subequations}\begin{eqnarray} \dot{\rho} &=& p_{\rho},\\ \dot{p}_{\rho} &=& \frac{1}{\Phi_g -E}\left[\frac{\partial\Phi_g}{\partial\rho}(1-p_{\rho}^2)-\frac{\partial\Phi_g}{\partial z}p_zp_{\rho}-\frac{L^2}{(E-\Phi_g)\rho^3}\right], \\ \dot{z} &=& p_z, \\ \dot{p}_z &=& \frac{1}{\Phi_g -E}\left[\frac{\partial\Phi_g}{\partial z}(1-p_z^2)-\frac{\partial\Phi_g}{\partial\rho}p_zp_{\rho}\right];\end{eqnarray} \label{RelGovern}\end{subequations} where $E$ is the constant value of energy related with the dynamical variables by the equation, \[p_{\rho}^2+p_z^2+\frac{L^2}{(E-\Phi_g)^2\rho^2}+\frac{1}{(E-\Phi_g)^2}=1.\] 

It is to be noted that in absence of the term representing the influence of halo (i.e. when $\alpha=0$), the problem becomes central, and hence integrable, ruling out any possibility of chaotic motion.

\section{Manifestation of the chaos}
Although the effective phase space is 4-dimensional, due to the autonomous nature of the governing equations,
eqs.~\eqref{govern}, the conservation of energy constraints the motion within a 3-dimensional constant energy hyper-surface in the 4-dimensional phase space, for a set of initial conditions. Now instead of observing the characteristic trajectories in a 3-dimensional space, it is customary, as well as convenient, to observe a 2-dimensional cross-section of that space. To generate that cross-section (known as Poincare section in the literature (e.g.~\cite{Berry1978}), 
one has to plot the points of intersection of the trajectories, moving along a particular direction (in or out), with a fixed plane within the phase space. As the trajectories are governed by the deterministic equations, the consecutive points of intersection are related by some deterministic map (Poincare map), but those relations may not be readily expressible using explicit analytic form.

For regular motion of an integrable system such a plot will be composed of systematic patterns; while chaotic motion produces a sea of scattered points 
(see, e.g., ~\cite{Berry1978,Goldstein2001} ). Generally a chaotic system does have islands of regular patterns surrounded the sea of scattered points (for soft chaos) and complete absence of any regular island of any length scale (hard chaos) rarely occur in practice. The extent of the sea of scattered points in comparison to such regular islands on the Poincare section gives a perceptible visual impression about the extent of chaos in the system concerned. 

The quantitative estimate of the degree of chaos may be evaluated by the average maximum Lyapunov exponent~\citep{Strogatz2007} of the phase trajectories. Lyapunov exponents are defined for a $D$-dimensional phase space (with co-ordinates $\bm{\mathit{x}}$) by  the eigenvalues of the matrix $\bm{\mathsf{M}}$ when the small deviations between neighbouring trajectories $\bm{\mathit{\delta x}}$ evolve with time (after linearisation) according to the equation \[\bm{\delta \dot{x}}=\bm{\mathsf{M}}\bm{\mathit{\delta x}}.\] But one may calculate a more readily computable quantity, \begin{equation}
\lambda= \lim_{t\rightarrow\infty}\lim_{\|\bm{\mathit{\delta x}}(0)\|\rightarrow 0}\frac{1}{t}\ln\left(\frac{\|\bm{\mathit{\delta x}}(t)\|}{\|\bm{\mathit{\delta x}}(0)\|}\right),  \label{LCN}                                                                                                                                                                                                                                                                                                                                                                                                                                                                                                                                                                                                                                                                                                                                                                                                                                                                                                                                                                                                                                                                                                                                                                                      \end{equation}
 where $\|\bm{\mathit{\delta x}}(0)\|$ is the initial norm of the deviation between two neighbouring points and  $\|\bm{\mathit{\delta x}}(t)\|$ is the same after evolving for time $t$ along their own trajectories, provided $\|\bm{\mathit{\delta x}}\|$ always remains much less than extent of allowable phase space; which tends to the largest Lyapunov exponent. Generally one calculates $\lambda$ several times for all the points spread over the allowable phase space uniformly and then take average over all such $\lambda$'s to find $\lambda_{\rm av}$. This $\lambda_{\rm av}$, called by some of the authors (e.g. \cite{Gueron2001,Ying2012,Froeschle1997}) as `Lyapunov characteristic number' (LCN), may be a rough estimate of the degree of chaos present in the system. In our work we made use of this $\lambda_{\rm av}$ as the quantitative indicator of the chaotic behaviour of the test particle.

\section{Numerical Results and Discussions}
\begin{figure*}[h!]
 \begin{tabular*}{1.0\linewidth}{ccc}
 \includegraphics[width=5cm]{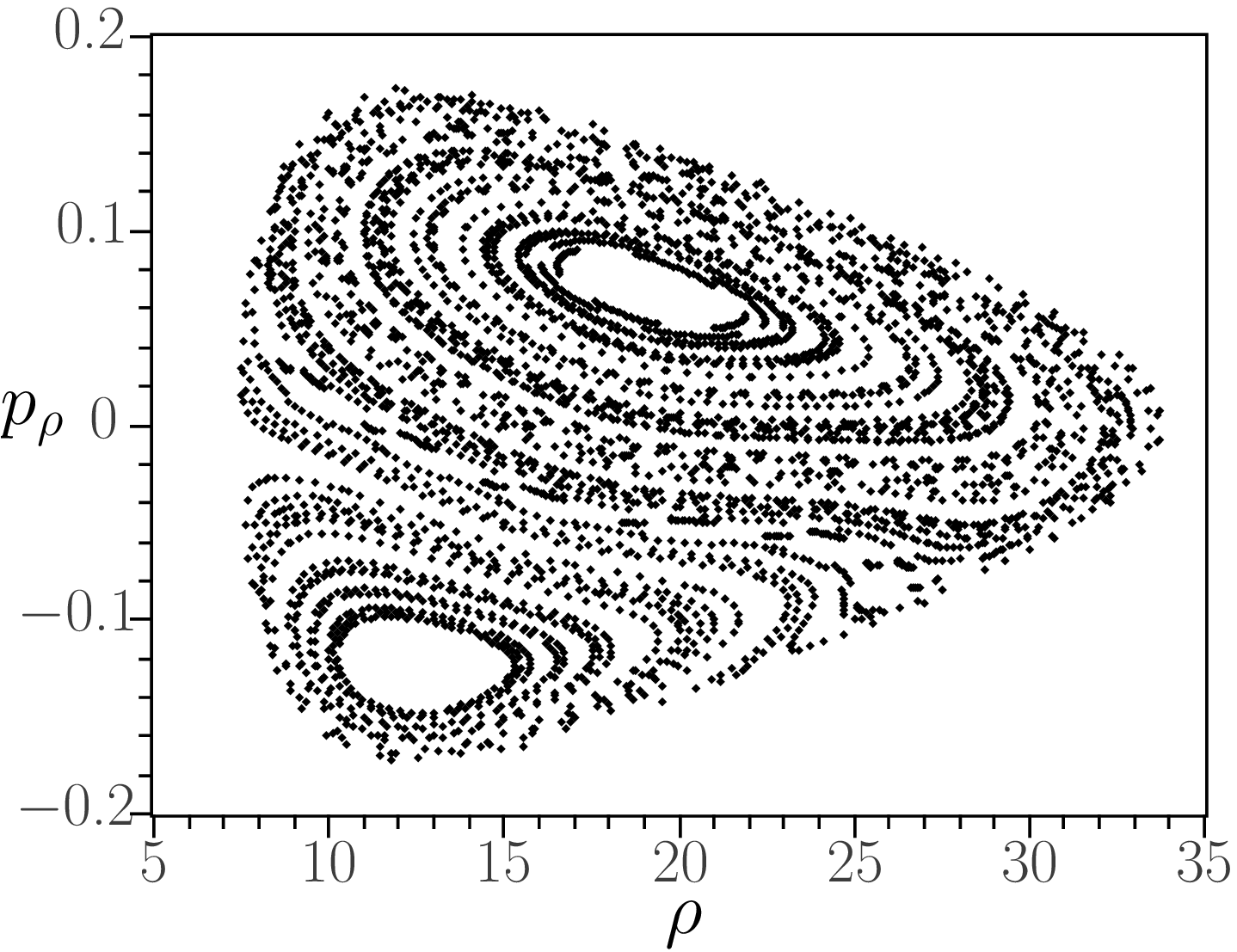} &
 \includegraphics[width=5cm]{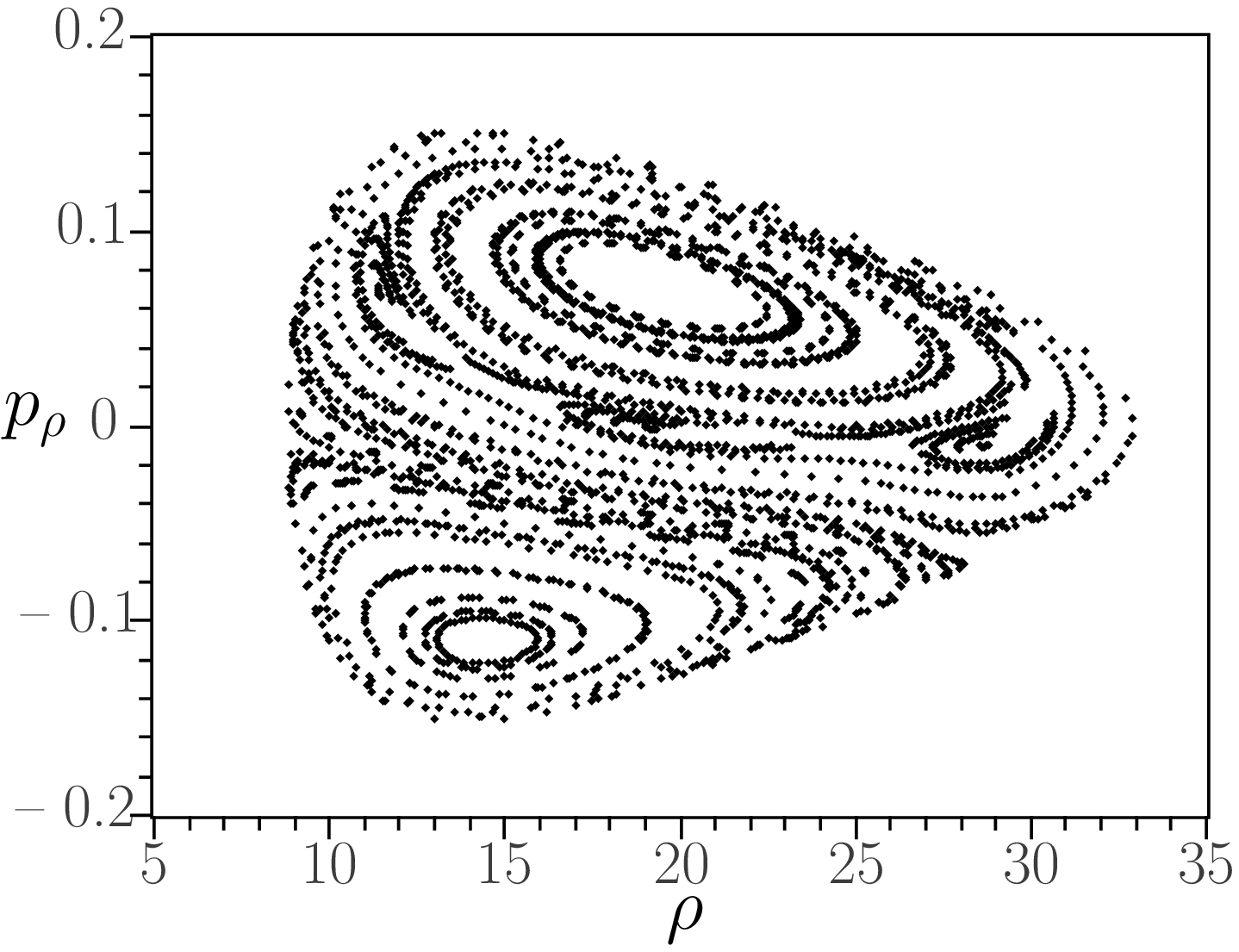} &
 \includegraphics[width=5cm]{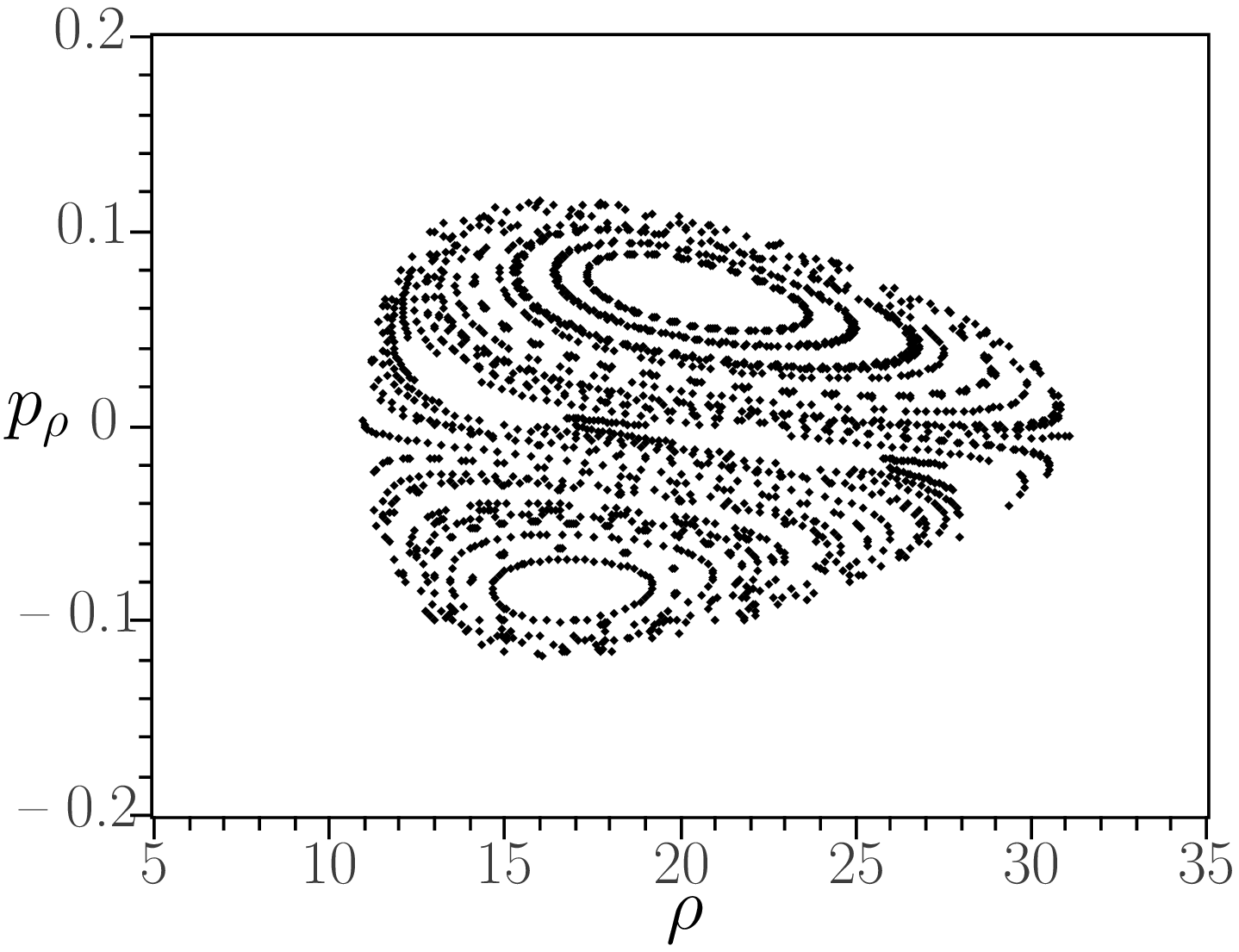}\\
 % fig11.eps: 0x0 pixel, 300dpi, 0.00x0.00 cm, bb=0 0 415 317
 (a)&(b)&(c)
 \end{tabular*}
\caption{\small Poincare section plots at the plane $z=0$ with $E=0.976$, $L=4.2$ and $\alpha=2\times 10^{-4}$, without any special relativistic correction. Variation of Kerr parameter: (a) $a=0.0$, (b) $a=0.3$ and (c) $a=0.8$.}
\label{fig1}
\end{figure*}
First of all we investigated the variation of the Poincare section taken at $z=0$, i.e. at equatorial plane, with variation of $E$, the energy of the test particle, other parameters viz. $L$ and $a$ of the system remaining fixed  (not shown here). It was observed qualitatively that the degree of chaos increases with increasing value of $E$, consistent with the earlier findings of some authors. This will be quantitatively shown later in this section.
\begin{figure*}
 \begin{tabular*}{1.0\linewidth}{ccc}
 \includegraphics[width=5cm]{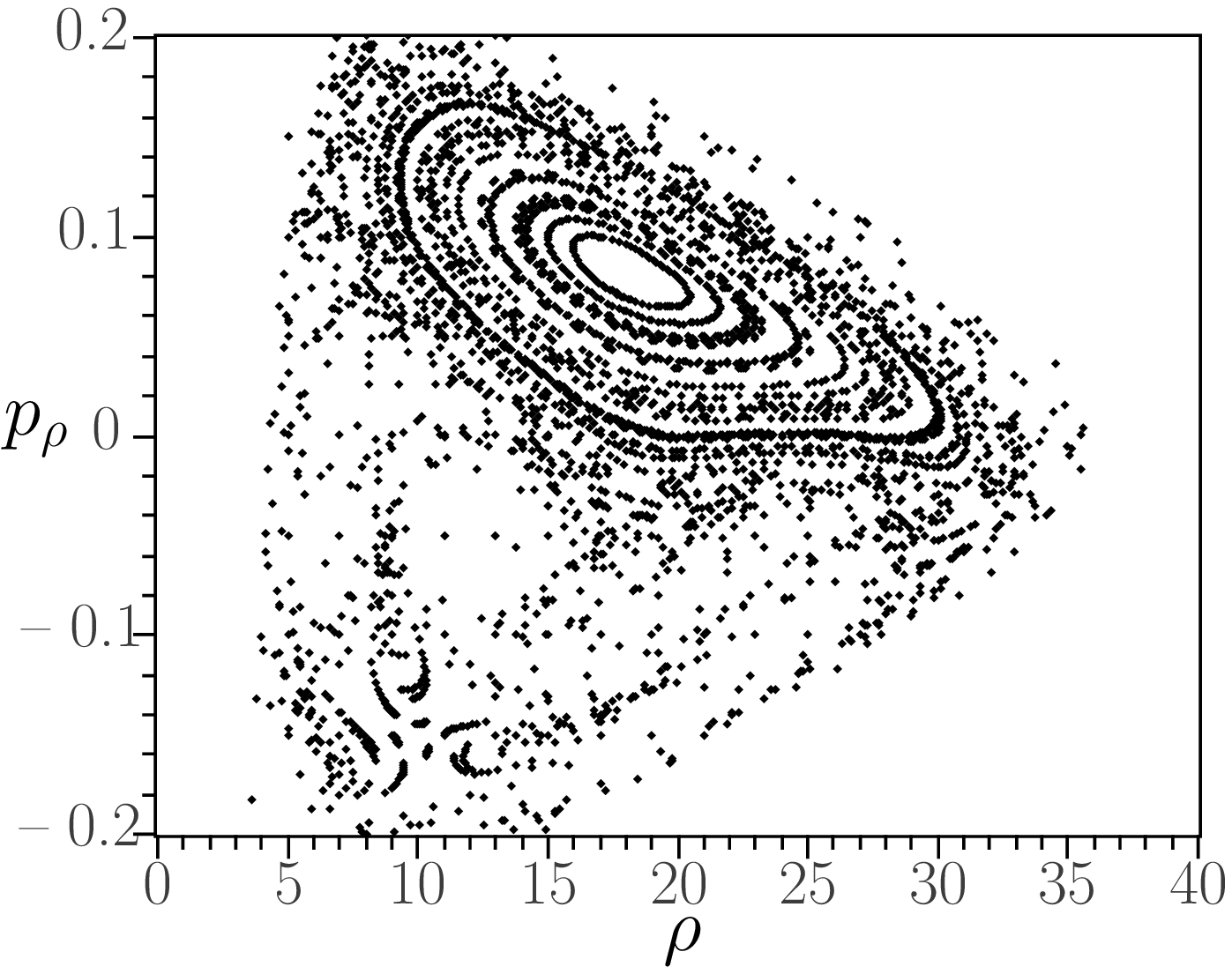} &
 \includegraphics[width=5cm]{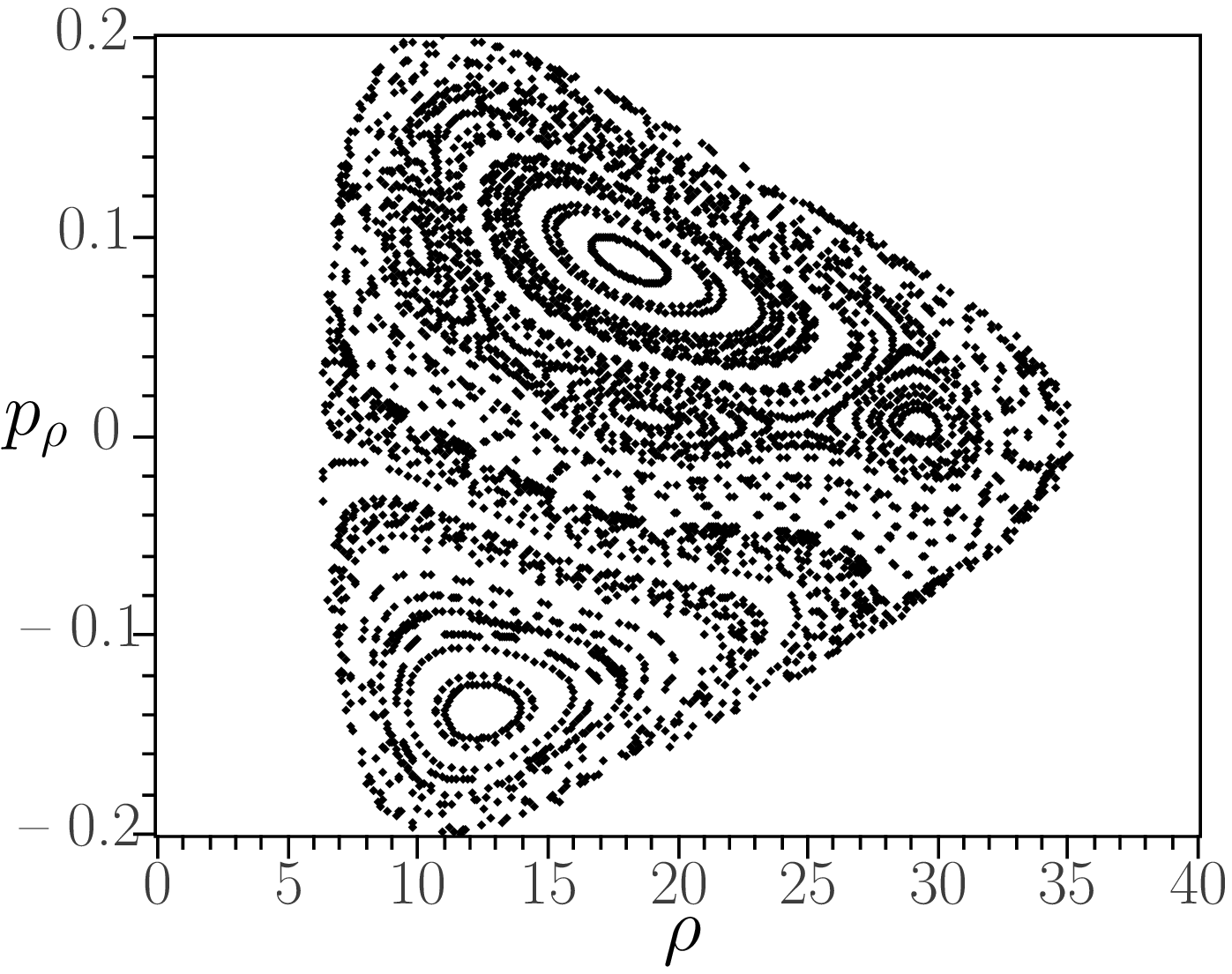} &
 \includegraphics[width=5cm]{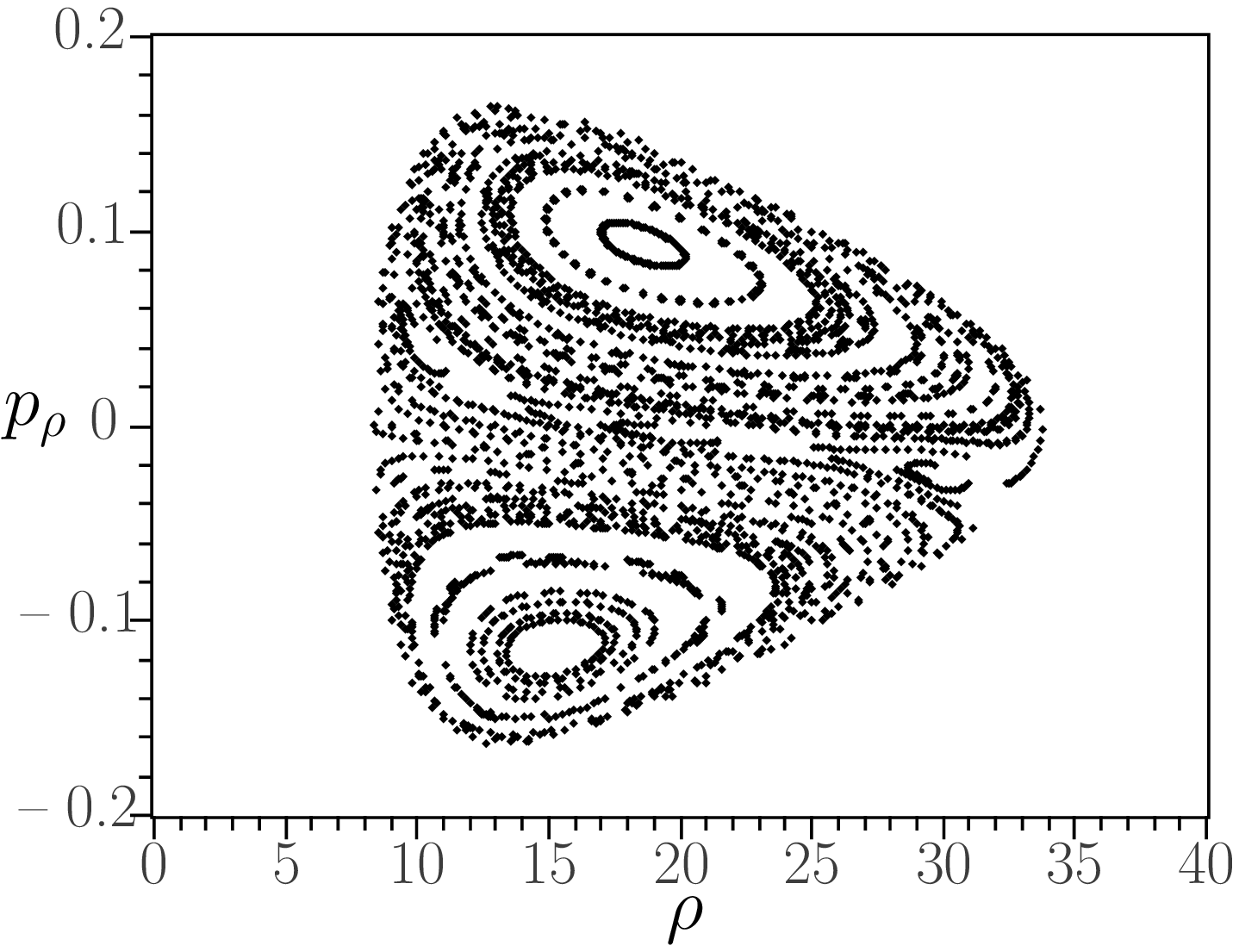}\\
 % fig11.eps: 0x0 pixel, 300dpi, 0.00x0.00 cm, bb=0 0 415 317
 (a)&(b)&(c)
 \end{tabular*}
\caption{\small Poincare section plots at the plane $z=0$ with $E=0.976$, $L=3.85$ and $\alpha=2\times 10^{-4}$, without any special relativistic correction. Variation of Kerr parameter: (a) $a=0.0$, (b) $a=0.3$ and (c) $a=0.8$.}
\label{fig2}
\end{figure*}

The variation of the degree of chaotic dynamics with that of $L$ is found (compare fig.~\ref{fig1} and~\ref{fig2}) to have a negative correlation, other parameters remaining same.

\begin{figure*}[h!]
 \begin{tabular*}{1.0\linewidth}{ccc}
 \includegraphics[width=5cm]{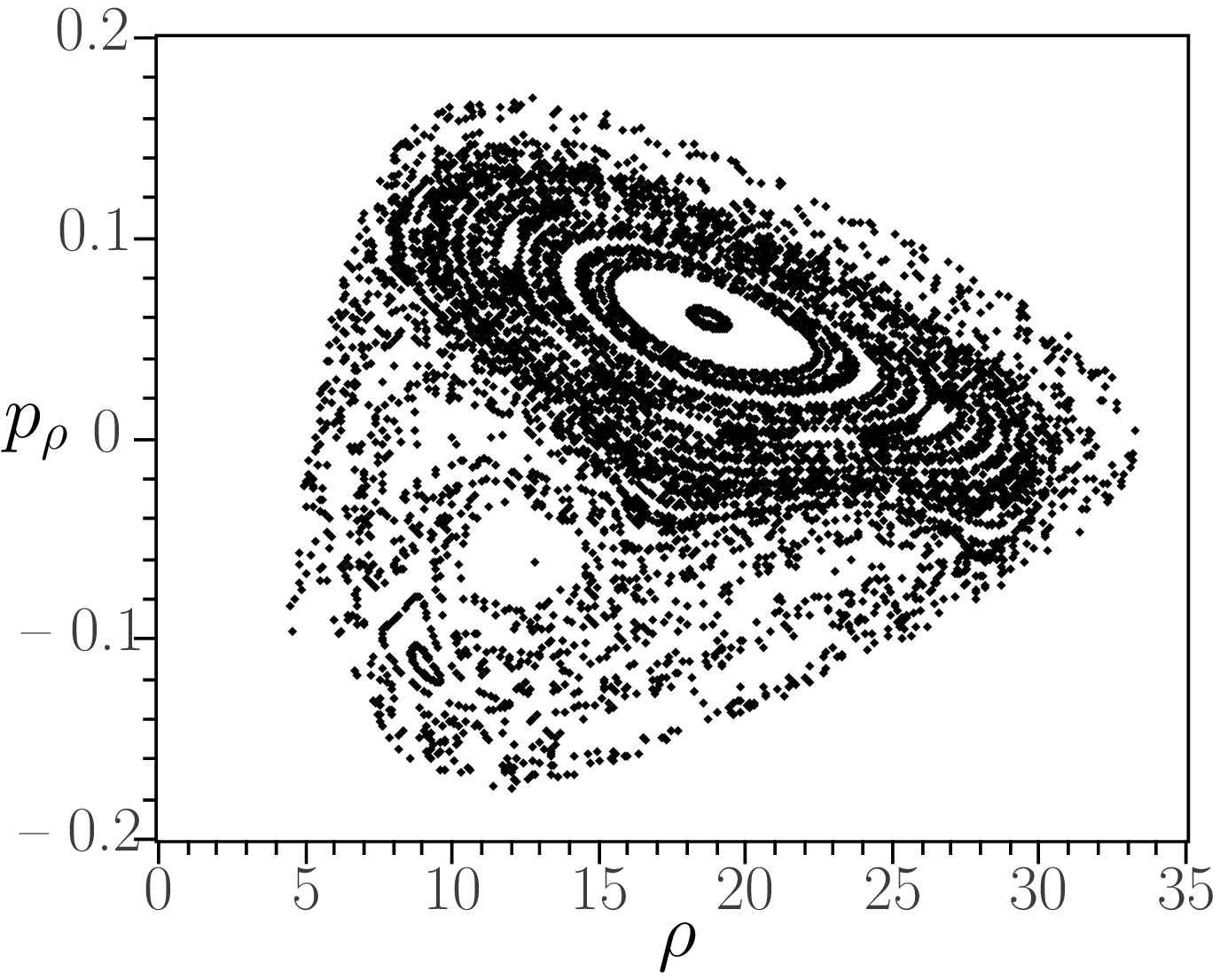} &
 \includegraphics[width=5cm]{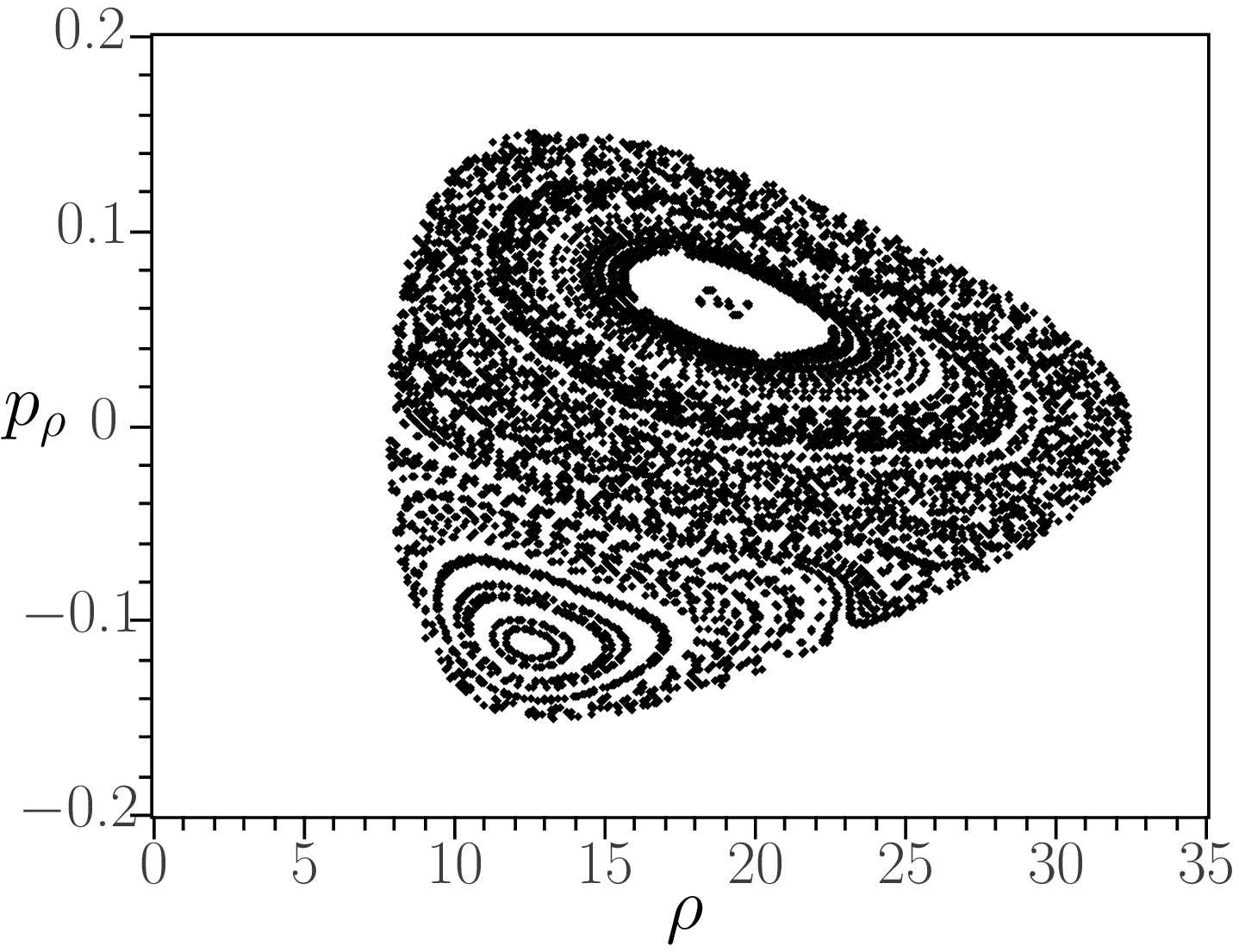} &
 \includegraphics[width=5cm]{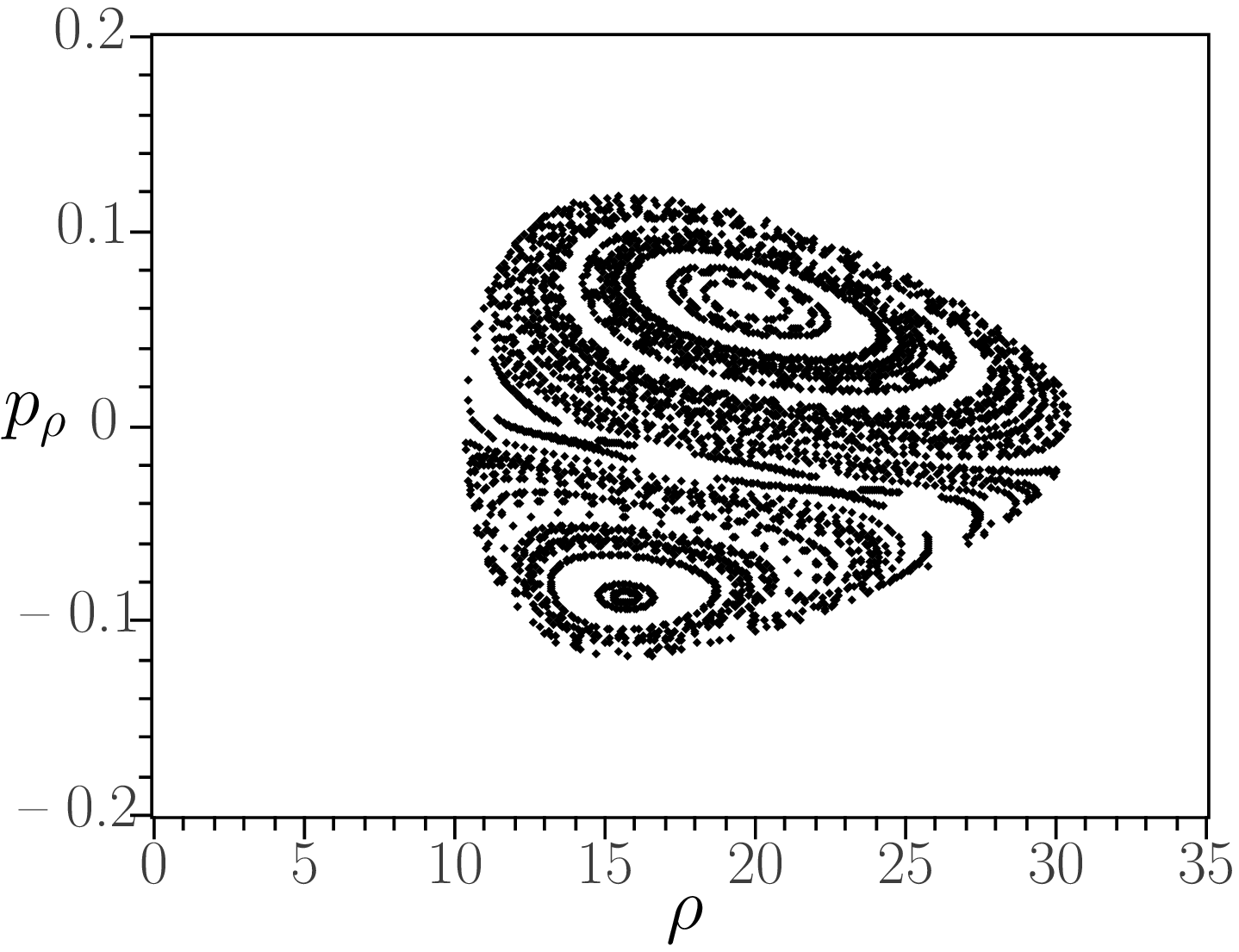}\\
 % fig11.eps: 0x0 pixel, 300dpi, 0.00x0.00 cm, bb=0 0 415 317
 (a)&(b)&(c)
 \end{tabular*}
\caption{\small Poincare section plots at the plane $z=0$ with $E=0.976$, $L=4.2$ and $\alpha=2\times 10^{-4}$, after special relativistic correction. Variation of Kerr parameter: (a) $a=0.0$, (b) $a=0.3$ and (c) $a=0.8$.}
\label{fig3}
\end{figure*}
The main focus of our study is to find the (anti)correlation between the black hole spin parameter and the degree of chaos of the test particle in the dipolar halo. It is apparent from figs~\ref{fig1} and~\ref{fig2} that with the increase of spin parameter $a$, the region covered by the scattered points decreases, indicating the reduction of chaotic behaviour of the motion of the test particle. To have a more accurate result one may be in interested to investigate the dynamics after special relativistic corrections using the equations. But the figs.~\ref{fig3} depicting the Poincare section after such corrections show more or less the same trend. However one feature that may be noticed by comparing figs.~\ref{fig2} and~\ref{fig3}, that the time evolution (as revealed by the Poincare section) after the correction roughly depicts the uncorrected motion with a somewhat lower value of $L$.

Further the study has been made to get a quantitative picture of the above correlation using the LCN. Though the values of LCN are rough estimates of the degree of chaos in the phase space, which in practice for numerical computation -- due to finite size of the sample drawn from the set of all allowable initial conditions (see equation~\eqref{LCN} -- always have an inherent fluctuation, the plot of LCN ($\lambda_{\rm av}$) over the range of $a$ shows a clear negative correlation, most prominent for the lower range of values of $a$.
\begin{figure*}[h]
 \centering
 \includegraphics[width=0.5\columnwidth]{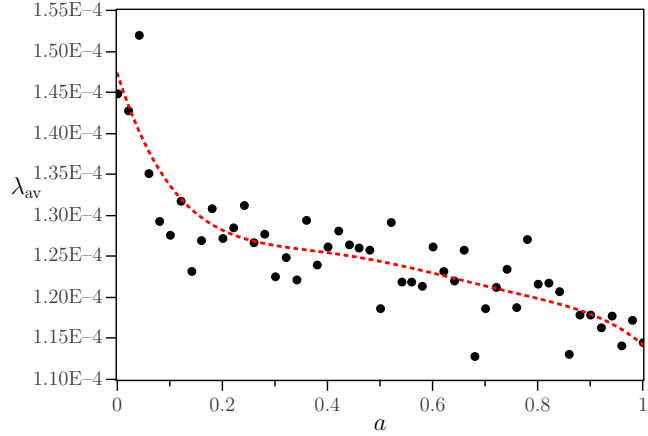}
 \caption{\small Variation of LCN ($\lambda_{\rm av}$) with Kerr parameter $a$ for $E=0.976$ and $L=3.85$ ($\alpha=2\times 10^{-4}$). The fitted curve shows a rough trend.}\label{fig4}
\end{figure*}

As the equations of motion without special relativistic corrections are to more or less mimic the motion after corrections, with a little higher value of $L$ and are yet much simpler to handle, we used the uncorrected version to find out the mechanism behind the suppression of chaos due to the increase of the black hole spin parameter $a$, observed numerically in this case. The investigation was done from the perspective of dynamical system analysis. 

\begin{figure*}[h!]
 \centering
 \includegraphics[width=0.4\columnwidth]{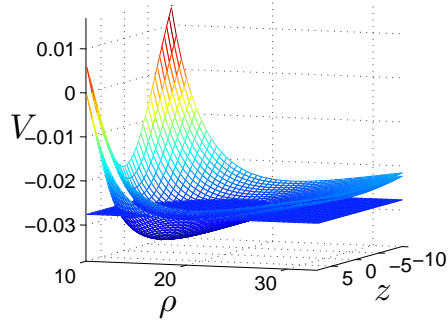}
 \caption{\small Constant effective potential (without the contribution from halo) surfaces for $a=0.3$ (lower one) and for $a=0.8$ upper one along $\rho$ and $z$. Here the parameter values, $L=4.2$. A constant energy plane is drawn to depict the allowed region of motion.}\label{fig5}
\end{figure*}

The potential $\Phi_{\rm eff}$ (excluding the perturbative term) is plotted with radial distance along the equatorial plane, i.e. for $z=0$, for different values of $a$ in the figure~\ref{fig5}  . It is apparent that $\Phi_{\rm eff}$ has a local minimum and the minimum value shifts upward as $a$ increases from $0$ to $1$, the shift being more prominent for lower values of $a$. With the help of non-relativistic equation of motion one may have following explanations for the suppression of chaos with the increase of black hole spin.

\begin{figure*}[h]
 \centering
 \includegraphics[width=0.4\columnwidth]{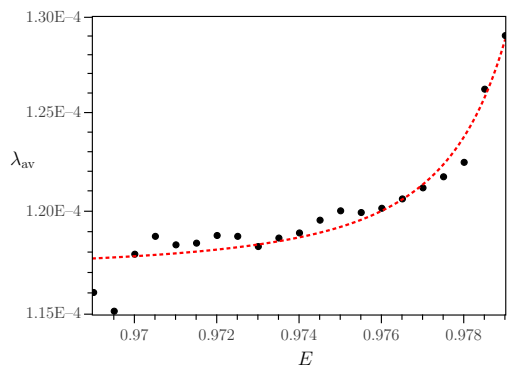}
 \caption{\small Variation of LCN ($\lambda_{\rm av}$) with the energy $E$ for $a=0.3$ and $L=3.85$ ($\alpha=2\times 10^{-4}$).}\label{fig6}
\end{figure*}

If the corresponding values of $\rho$ and $z$ at the minimum are denoted by $\rho_0\;\textrm{and}\; z_0$ respectively, the potential at the neighbourhood of that point may be approximated by \[V\approx V(\rho_0,z_0)+\left.\frac{\partial^2 V}{\partial\rho^2}\right|_0 (\rho-\rho_0)^2+\left.\frac{\partial^2 V}{\partial z^2}\right|_0 (z-z_0)^2.\] Thus it is to be noticed that if one expresses the equations~\eqref{govern}, in terms of this approximated effective potential, the equations, even after incorporating $\alpha z$ term in the potential (which may only shift $\rho_0$, $z_0$ slightly), become linear, ruling out any chaos. So as long as the motion is restricted near the local minimum only, i.e. for the low values of energy, it will be near integrable. As the energy becomes lower the motion becomes more restricted near the point with minimum potential as the from the energy consideration, the motion is allowed only within the region where $E_{\rm mech}-\Phi>0$ ($E_{\rm mech}=(E^2-1)/2$ being the mechanical energy of the test particle excluding its rest mass energy). The argument is further confirmed by the trend of the LCN in the fig.~\ref{fig6}.

Again on the other hand, for a particular value energy $E$, with the increase of $a$ (as well as of $L$, not shown) the minimum value of the effective potential $V(\rho_0,z_0)$ increases (see fig.~\ref{fig5}), making the motion more restricted near the minimum, keeping less room for the governing equations being non-linear with chaotic solutions. This explains the suppression of chaos with the increase of $a$ (as well as with the increase of $L$, already manifested qualitatively).

%%%%%%%%%%%%%%%%%%%%%%%%%%%%%%%%%%%%%%%%%%%%%%%%%%%%%%%%%%%%%%%%%%%%%%%%%%%%%%%%%%%%%%%%%%%%%%%%%%%%%%%%%

\section*{Acknowledgements}

The work of DA has been done during her long term stay at Harish Chandra 
Research Institute (HRI) as a project assistant (position funded by the Cosmology \& High 
Energy Astrophysics planned project fund under the XII th plan at HRI).    
SN and  SS would like to acknowledge the kind hospitality provided by HRI, Allahabad,
India, for multiple visits. The work of SN has partially been
supported by the UGC MRP grant (sanction no: F. PSW -- 163/13-14). 

%\bibliography{halochaos}
%\bibliographystyle{plainnat.bst}

\end{document}